\documentclass[journal,12pt,onecolumn,draftclsnofoot]{IEEEtran}
\IEEEoverridecommandlockouts
\usepackage{cite}
\usepackage{amsmath,amssymb,amsfonts}
\usepackage{algorithmic}
\usepackage{graphicx}
\usepackage{subcaption}
\usepackage{url}
\usepackage{bbm}
\usepackage{bm}
\usepackage{siunitx}
\usepackage{mathtools, nccmath}
\usepackage{textcomp}
\usepackage{xcolor}
\usepackage{mathtools}
\usepackage{blindtext}
\usepackage[utf8]{inputenc}
\usepackage{float}
\usepackage{enumitem}
\usepackage{hyperref}

\newcommand{\refappendix}[1]{\hyperref[#1]{Appendix~\ref*{#1}}}
\usepackage{graphicx}
\usepackage{multirow}
\usepackage{tabularx}

\usepackage{acronym}
\acrodef{sc}[SC]{small cell}
\acrodef{ppp}[PPP]{Poisson point process}
\acrodef{ul}[UL]{uplink}
\acrodef{bs}[BS]{base station}
\acrodef{snr}[SNR]{signal-to-noise ratio}
\acrodef{ue}[UE]{user equipment}
\acrodef{pdf}[PDF]{probability density function}
\acrodef{los}[LOS]{line of sight}
\acrodef{nlos}[NLOS]{non-line of sight}
\acrodef{pgfl}[PGFL]{probability generating functional}
\acrodef{cdf}[CDF]{cumulative density function}
\acrodef{pdf}[PDF]{probability density function}
\acrodef{mmw}[mmWave]{millimeter waves}
\acrodef{mimo}[MIMO]{multiple-input, multiple-output}

\def\BibTeX{{\rm B\kern-.05em{\sc i\kern-.025em b}\kern-.08em
    T\kern-.1667em\lower.7ex\hbox{E}\kern-.125emX}}
\begin{document}
\title{\LARGE{Effect of Correlated Building Blockages on the Ergodic Capacity of mmWave Systems in Urban Scenarios}\\
\thanks{

Affiliation: Dept. Signal Theory and Communications, Universitat Politècnica de Catalunya, Barcelona, Spain. Emails: \IEEEauthorrefmark{1}cristian.garcia.ruiz@estudiantat.upc.edu,
\IEEEauthorrefmark{2}olga.munoz@upc.edu,
\IEEEauthorrefmark{3}antonio.pascual@upc.edu.

The work presented in this paper has been funded through the project ROUTE56 - PID2019-104945GB-I00 (funded by Agencia Estatal de Investigación, Ministerio de Ciencia e Innovación, MCIN / AEI / 10.13039/501100011033).

\copyright 2022 IEEE. Personal use of this material is permitted. Permission from IEEE must be obtained for all other uses, in any current or future media, including reprinting/republishing this material for advertising or promotional purposes, creating new collective works, for resale or redistribution to servers or lists, or reuse of any copyrighted component of this work in other works.

DOI: 10.1109/TVT.2022.3155356

}
}

\author{\IEEEauthorblockN{Cristian García Ruiz\IEEEauthorrefmark{1},
Olga Muñoz\IEEEauthorrefmark{2}, \IEEEmembership{Member, IEEE}},
Antonio Pascual-Iserte\IEEEauthorrefmark{3}, \IEEEmembership{Senior Member, IEEE}\vspace{-0.9cm}}


\markboth{Accepted paper at IEEE Transactions on Vehicular Technology (Volume: 71, Issue: 5, May 2022)}
{}

\maketitle

\begin{abstract} The \ac{mmw} bands, considered to support the forthcoming generation of mobile communications technologies, have a well-known vulnerability to blockages. Recent works in the literature analyze the blockage probability considering independence or correlation among the blocking elements of the different links. In this letter, we characterize the effect of blockages and their correlation on the ergodic capacity. We carry out the analysis for urban scenarios, where the considered blocking elements are buildings that are primarily parallel to the streets. We also present numerical simulations based on actual building features of the city of Chicago to validate the obtained expressions.
\end{abstract}


\newpage
\section{Introduction}
\subsection{Background and Motivation}
\IEEEPARstart{T}{he} problem of blockages in \ac{mmw} and the computation of relevant metrics such as blockage probability has received considerable attention in the recent literature \cite{bai2013,hriba:2019,Gupta2020DoesBC,garcia2021,Article_correlation}, as an obstacle in the middle of a  \ac{mmw} link can impede the communication between a user and the serving \ac{bs} completely. Assuming that users can only communicate with a visible \ac{bs}, that is, in \ac{los}, an interesting problem is to relate the blockages in the paths to different \acp{bs} with the achievable communication rate. In relation to this problem, the goal of this paper is to provide an approximation of the \ac{ul} ergodic capacity in a scenario of \acp{bs} with random positions, as modeled in \cite{bai2013}.

The addition of blockages in the scenario yields us to the central
problem of the derivation of the distribution of the distance
to the closest visible BS when a set of elements may obstruct
the communication links. In \ac{mmw}, communication blockages can be caused by large static objects such as buildings in urban scenarios \cite{bai2013,Gupta2020DoesBC,garcia2020,Article_correlation,garcia2021,ITU-R_P.1410-5}, or small dynamic objects such as human body blockages \cite{Article_correlation,Samuylov2016,jain2019TIM}, both types resulting in \ac{nlos} situations. In this paper, we consider the first type of blockages, that is, blockages due to buildings in urban scenarios. We will leave for future work the effect of human body blockage, which will translate into an increase of the total blockage probability.

Although initial works on blocking characterization did not consider the correlation among blockages \cite{bai2013}, its impact on the blockage probability was studied  in \cite{hriba:2019,Article_correlation,Conference_correlation,garcia2020,garcia2021}. The bounds for the distribution of the distance to the closest visible \ac{bs} derived in \cite{Gupta2020DoesBC} also considered the effect of correlation. All these works showed that correlation has a relevant impact and can not be neglected. Although in \cite{Gupta2020DoesBC} blockages with random orientations were considered, in urban scenarios, buildings are regularly deployed, forming streets. Therefore, the impact of this regularity must also be considered. Accordingly, in this paper, we include this pattern in the blockages model to compute the \ac{cdf} of the distance to the closest visible \ac{bs}. Then, we use this result to study the effect of correlated building blockages on the ergodic capacity.

\vspace{-0.3cm}
\subsection{Goals and Contributions}
The contributions of this work are, for urban scenarios with blocking elements parallel to the streets, the derivation of:
\begin{itemize}
    \item the \ac{cdf} of the distance to the closest visible \ac{bs} for both uncorrelated and correlated building blockages,
    \item a methodology to consider the heights of the blockages,
    \item an approximation of the \ac{cdf} of the ergodic capacity.
\end{itemize}

\vspace{-0.2cm}
\section{System Model} \label{sec:System_Model}
\begin{figure}[t]
	\centering
	\includegraphics[width=0.8\columnwidth]{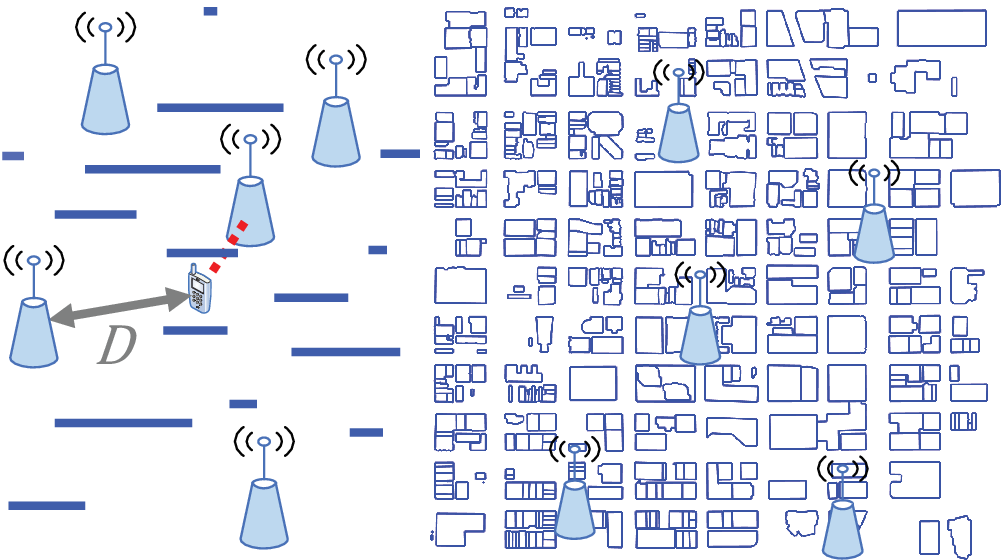}
	\caption{Scenario with blockages with parallel orientations (left) and a real deployment in Chicago (right).\vspace{-0.3cm}}
	\label{fig:Scenario_with_blockages_parallel_orientation_real_deployment_Chicago}
\end{figure}
We consider a scenario with a set of \acp{bs} (such as small cells) whose positions follow a \ac{ppp} $\Psi_{\text{BS}}$ with uniform density $\lambda_{\text{BS}} \hspace{1mm}\SI{}{BSs/m^2}$. We assume that intracell users share orthogonal resources while adjacent \acp{bs} are coordinated by employing different frequencies, leading to an interference-free scenario (as assumed in \cite{calvo2012,aravanis2016}).

Following \cite{randomshape,bai2013,garcia2020}, we model the blocking elements as line segments with random lengths represented by the random variable (r.v.) $L$ with \ac{pdf} $f_L(l)$. The centers of the segments follow a \ac{ppp} $\Psi_{\text{b}}$ with density $\lambda_{\text{b}} \hspace{1mm}\SI{}{blockages/m^2}$. Inspired by the regularity of buildings in cities, we consider parallel segments. Although this simple model does not account for the thickness of the buildings, the addition of a correction factor to the density can capture this effect, as shown in \cite{garcia2020} and later in this paper.

Given the uniformity of the scenario, we consider, without loss of generality, a reference user at the origin \cite{randomshape,bai2013}. The user is served only by the closest visible \ac{bs}, that is, in \ac{los}.

As the heights of in-street users are low, we assume that any building in the line connecting the user and a \ac{bs} will cause \ac{nlos} regardless of the height of such building. This assumption reduces the complexity of the analysis.

An example of this scenario is shown on the left in \autoref{fig:Scenario_with_blockages_parallel_orientation_real_deployment_Chicago}. On the right, a real layout of the Chicago city can be seen \cite{measurementsKulkarni}.

In the \ac{ul}, the \ac{snr} $\gamma$ at $\SI{1}{\meter}$ is
\begin{align}\label{eq:gamma}
    \gamma = \frac{P_{\text{U}}}{\sigma^2L_{\text{ref}}},
\end{align}
where $P_{\text{U}}$, $\sigma^2$, and $L_{\text{ref}}$ account for the user power, the noise power, and the equivalent path-loss at a distance of $\SI{1}{\meter}$, respectively. Transmitter and receiver antenna gains and multi-antenna processing gains could be included in \eqref{eq:gamma} as in \cite{aravanis2016,calvo2012}. Similarly, another term could be added to account for other losses due to, for example, atmospheric effects \cite{garcia2021}. However, these terms are omitted for simplicity in the notation.

\vspace{-0.15cm}
\section{Bound on the Ergodic Capacity} \label{sec:Bound_Ergodic_Capacity}

Let $D$ and $R$ be the r.v.s (with \acp{pdf} $f_D(d)$, $f_R(r)$; and \acp{cdf} $F_D(d)$, $F_R(r)$) representing the distance between the reference user and the closest visible BS and the corresponding UL ergodic capacity. The r.v.s $D$ and $R$ depend on the statistics of the blockages (e.g., density, correlation among them, etc.) and are related to each other as follows: 
\begin{align} \label{eq:ergodic_capacity}
    R=\mathbb{E}_{h}\left[\log \left( 1+ D^{-\alpha}{\lvert h \rvert}^2\gamma \right)\right],
\end{align}
where $h$ denotes the fast fading Rayleigh channel that follows a complex zero-mean circularly symmetric Gaussian distribution with a variance equal to 1, $\alpha$ is the path-loss exponent, and $\gamma$ is the UL SNR at 1 m defined in eq. (\ref{eq:gamma}). Whenever there is no visible BS (i.e., all the \acp{bs} are blocked), the r.v.s $D$ and $R$ become infinity and 0, respectively.

A lower bound of the ergodic capacity in eq. (\ref{eq:ergodic_capacity}) was given in \cite{calvo2009}. This bound, which we will denote with the r.v. $\underline{R}$, is
\begin{align} \label{eq:rate_lb}
   \underline{R} \hspace{-0.2mm}  = \hspace{-0.2mm} \log \hspace{-0.5mm} \left(\hspace{-0.5mm}1+ \gamma \rho e^{-\alpha \log (D)} \hspace{-1mm}\right) \hspace{-0.2mm} \leq \hspace{-0.2mm} \mathbb{E}_{h} \hspace{-0.5mm}\left[\log \hspace{-0.5mm} \left ( \hspace{-0.5mm} 1+ D^{-\alpha}{\lvert h \rvert}^2\gamma \hspace{-0.5mm} \right)\right],
\end{align}
where the inequality comes from Jensen's inequality and the convexity of the $\log (1+e^x)$ function, and $\rho$ depends on the expectation of the logarithm of a Chi-square r.v. as follows
\begin{align}
    \rho = e^{\mathbb{E}_h\left[\log({\lvert h \rvert}^2) \right]} = e^{-\chi},
\end{align}
with $\chi \approx 0.577$ being the Euler-Mascheroni constant \cite{calvo2009}.

Our goal is to find an approximation of the \ac{cdf} of the ergodic capacity for the case of parallel blockings elements. To that end, we derive the \ac{cdf} of the lower bound of the rate, $F_{\underline{R}}(r)$, through the \ac{cdf} $F_D(d)$. According to \eqref{eq:rate_lb}, we have that $D$ and $\underline{R}$ are related through the function $v(\cdot)$ as follows:
\begin{eqnarray}
    \underline{R} & = & v(D) = \log \left(1 + \gamma \rho e^{-\alpha \log (D)} \right)\\
    D & = & v^{-1}(\underline{R}) = {\left(\frac{\gamma \rho}{e^{\underline{R}}-1}\right)}^{\frac{1}{\alpha}}.
\end{eqnarray}

Given that $v(D)$ is monotonically decreasing, we have that:

\begin{equation} \label{eq:F_R(r)}
F_{\underline{R}}(r) = 1 - F_D\left(v^{-1}(r)\right) = 1 - F_D\left({\left(\frac{\gamma\rho}{e^{r}-1}\right)}^{\frac{1}{\alpha}}\right).
\end{equation}
Note that $F_R(\cdot)$ is upper bounded by $F_{\underline{R}}(\cdot)$ since
\begin{equation}
F_R(r) = \mathbb{P}(R \leq r) \leq \mathbb{P}(\underline{R} \leq r) = F_{\underline{R}}(r).
\end{equation}
Accordingly, we can guarantee that users will achieve at least a rate $r$ with a probability $1-F_R(r)$ greater than $1-F_{\underline{R}}(r)$.

\vspace{-0.2cm}
\section{Distribution of the Distance to the Closest Visible BS} \label{sec:Distribution_Distance_Closest_Visible_BS}
The objective of this section is to obtain the \ac{cdf} $F_D(d)$. After a brief introduction of the notation, we consider the cases of no blockages \cite{bai2013}, independent blockages, and correlated blockages in subsections IV-A, IV-B, and IV-C, respectively.

Let $\mathcal{B}\left (\text{\textbf{O}},d\right)$ denote a ball in $\mathbb{R}^2$ of radius $d$ centered at the origin \textbf{O} and $P$ the number of \acp{bs} (in  \ac{los} or \ac{nlos}) falling inside the ball.
Let us also define the event $T_p$ as in \cite{bai2013}:
\begin{equation} \label{eq:Event_Tp}
T_p = \left\{\text{there are }P=p \text{ BSs in }\mathcal{B}\left (\text{\textbf{O}},d\right) \right\}.
\end{equation}

Since the \acp{bs} follow the \ac{ppp} $\Psi_{\text{BS}}$, then $P \sim \mathcal{P}\left(\lambda_{\text{BS}} \pi d^2 \right)$, being $\mathcal{P}(\chi)$ a Poisson distribution of parameter $\chi$ \cite{bai2013}.

\vspace{-0.3cm}
\subsection{Blockage Free Scenario}\label{subsec:Without_Blockages}
In the absence of blockages, all the \acp{bs} are in \ac{los}. Therefore, $\mathbb{P}\left(D > d \right) = \mathbb{P}\left(T_0 \right)$, i.e., the probability that zero \acp{bs} fall within $\mathcal{B}\left (\text{\textbf{O}},d\right)$, which leads to \cite{bai2013, aravanis2016}:
\begin{align} \label{eq:F_D_without_blockages}
    F_D(d) = \mathbb{P}\left(D \leq d \right) = 1- \mathbb{P}\left(T_0 \right) = 1 - e^{-\lambda_{\text{BS}} \pi d^2}.
\end{align}

\subsection{Scenario with Blockages: Independent Case} \label{subsec:Independent_Case}
\begin{figure}[t]
	\centering
	\includegraphics[trim=0 0 0 0, clip, width=5cm]{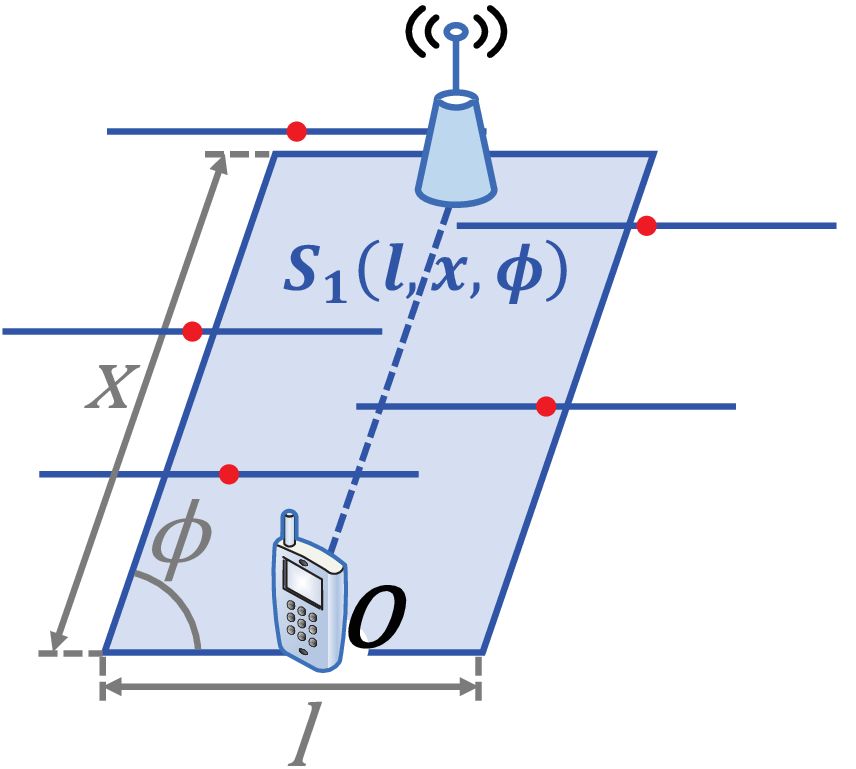}\vspace{-0.1cm}
	\caption{Blocking region $S_1\left(l, x, \phi\right)$ of a single link.\vspace{-0.5cm}}
	\label{fig:Blocking_region_single}
\end{figure}

When considering blockages, the user can only connect to the closest \ac{bs} in \ac{los}. Following \cite{bai2013}, let us define the event
\begin{equation} \label{eq:Event_Z}
	Z = \left\{\text{no BS in }\mathcal{B}\left (\text{\textbf{O}},d\right) \text{ is visible to \textbf{O}} \right\}.
\end{equation}
Since a \ac{bs} can be in \ac{nlos}, the probability that the distance to the closest visible \ac{bs} $D$ is greater than  $d$ is
\begin{align} \label{eq:P_Z_general_parallel_orientation}
    \mathbb{P}\left(D > d \right) = \mathbb{P}\left(Z \right) = \sum_{p=0}^{\infty} \mathbb{P}\left(Z | T_p\right) \mathbb{P}\left(T_p \right).
\end{align}
Assuming independence of the blockages in the different links, we have that
\begin{equation} \label{eq:P_Z_T_p_independent_parallel_orientation}
    \mathbb{P}\left(Z | T_p \right) = {\left(\mathbb{P}\left(Z | T_1 \right) \right)}^p.
\end{equation}
Despite this simplification, obtaining $\mathbb{P}\left(Z | T_1 \right)$ is not straightforward since the event $T_1$ only gives information about the occurrence of a \ac{bs} falling within $\mathcal{B}\left (\text{\textbf{O}},d\right)$, but not about its position. Taking $X$ and  $\Phi$ as the distance and the azimuth of the only \ac{bs} within the ball with respect to $\textbf{O}$, we have that:
\begin{align}
    \mathbb{P}\left(Z | T_1 \right) = \mathbb{E}_{x,\phi} \left[\mathbb{P}\left(Z | T_1, X=x, \Phi=\phi \right)\right],
\end{align}
that is, to obtain this probability, we need to average with respect to all the possible positions of the \ac{bs} inside the ball. Now, unlike \cite{bai2013}, the average on $\phi$ can not be omitted when the blocking elements are parallel as in urban scenarios, which makes the derivation quite more challenging due to the lack of circular symmetry. To derive the expression for $\mathbb{P}\left(Z | T_1, X=x, \Phi=\phi \right)$, we first assume that blockages have a given length $l$ and that the \ac{bs} is located at an azimuth $\phi$ and a distance $x$ with respect to the reference user, as depicted in \autoref{fig:Blocking_region_single}. In the figure, the parallel lines (in blue, with centers in red) are examples of blockages that effectively block this link 1. The shadowed parallelogram $S_1\left(l, x, \phi\right)$, called blocking region, is the region where the center of a blocking object should lie to block the link. As seen in \cite{bai2013,garcia2020}, the area of this parallelogram is $A_{S_1}\left(l, x, \phi\right) = x \cdot l \lvert \sin{\left(\phi \right)} \rvert $. Accordingly, the number of blockages of length $l$ blocking link 1, between this \ac{bs} and the reference user, is denoted by $K_{1_l}$, where $K_{1_l} \sim \mathcal{P}\left(\lambda_{b_l} A_{S_1}\left(l, x, \phi\right) \right)$, with $\lambda_{b_l} = \lambda_{\text{b}} f_L(l) \text{d}l$ and $\mathbb{E}\left[K_{1_l}\right]=\lambda_{b_l} A_{S_1}(l, x, \phi)$ \cite{bai2013,Article_correlation}.
Consequently, the mean number of blockages with any length, denoted by $K_1$, is
\begin{align}
    \mathbb{E}\left[K_1\right] = \int_l \mathbb{E}\left[K_{1_l}\right] = \lambda_{\text{b}} \mathbb{E}\left[L\right] x \lvert \sin{\left(\phi \right)} \rvert,
\end{align}
which yields us to: 
\begin{align} \label{eq:P_NLOS_parallel_orientation}
    \mathbb{P}\left(Z | T_1, X=x, \Phi=\phi \right) = 1 - e^{-\mathbb{E}\left[K_1\right]}.
\end{align}

The consideration of the uniformity of the positions of the \ac{bs} within $\mathcal{B}\left (\text{\textbf{O}},d\right)$ implies that $f_X(x) = \frac{2x}{d^2}$ with $x \in [0,d]$ and $f_{\Phi}(\phi) = \frac{1}{2\pi}$ with $\phi \in [0,2\pi]$ \cite{bai2013}. Therefore,
\begin{eqnarray}
    && \hspace{-6.5mm} \mathbb{P}\left(Z | T_1 \right) =  \mathbb{E}_{x,\phi} \left[\mathbb{P}\left(Z | T_1, X=x, \Phi=\phi\right)\right] \nonumber \\
    &\overset{\text{(a)}}{=}& 1 - \frac{4}{\pi d^2} \int_0^{\frac{\pi}{2}} \int_0^d x e^{-\lambda_{\text{b}} \mathbb{E}\left[L\right] x \sin{\left(\phi \right)}} \text{d}x \text{d}\phi \nonumber \\
    &\overset{\text{(b)}}{\approx}& 1 - \frac{4}{\pi{\left(\lambda_{\text{b}} \mathbb{E}\left[L\right] d \right)}^2 m n \left(n + \frac{m\pi}{2}\right) } \times \label{eq:P_Z_T_1_parallel_orientation} \\
    &&\left(\frac{m\pi}{2} + n e^{-\lambda_{\text{b}} \mathbb{E}\left[L\right] d \left(n + \frac{m\pi}{2}\right)} - \left(n  +  \frac{m\pi}{2}\right) e^{-\lambda_{\text{b}} \mathbb{E}\left[L\right] d n} \right), \nonumber
\end{eqnarray}
where (a) comes from the symmetry of $\lvert \sin{\left(\phi \right)} \rvert$ and (b) from the linear approximation of $\sin{\left(\phi \right)}$ for $\phi \in \left[0, \frac{\pi}{2}\right]$ with $\sin{\left(\phi \right)} \approx m\phi + n$ with $m = \frac{96\pi-24}{4\pi^4-3\pi^2} = 0.7710$ and $n = \frac{8 - m\pi^2}{4\pi} = 0.0311$ ($m$ and $n$ are obtained through the minimization of the square error of this approximation).

Finally, using \eqref{eq:P_Z_general_parallel_orientation}, the upper bound of $F_D(d)$ can be obtained assuming independence among the blockages as
\begin{equation}\label{eq:F_D(d)_independent}
    \overline{F_D(d)} = 1 -\underline{\mathbb{P}\left(D > d \right)} = 1 - e^{-\lambda_{\text{BS}} \pi d^2 \left(1-\mathbb{P}\left(Z | T_1 \right)  \right)},
\end{equation}
where $\mathbb{P}\left(Z | T_1 \right)$ is given by \eqref{eq:P_Z_T_1_parallel_orientation} and the formulation of $\underline{\mathbb{P}\left(D > d \right)}$  in \cite{bai2013} (included for convenience in \refappendix{subsecapp:Derivation_of_P_D(d)_lb}).

\vspace{-0.2cm}
\subsection{Scenario with Blockages: Correlated Case} \label{subsec:Correlated_Case}
In this subsection, following \cite{Gupta2020DoesBC}, only the correlation between the reference link (namely link 1) and the $i$-th link (for $i \geq 2$) is considered, one by one, while the correlation among more than two links at the same time is not taken into account.

We compute first $\mathbb{P}\left(D > d \right) = \mathbb{P}\left(Z \right)$, which is the probability that either the closest visible \ac{bs}, located at \textbf{x}, is outside $\mathcal{B}\left (\text{\textbf{O}},d\right)$ or that there is no visible \ac{bs}. The latter is equivalent to the fact that the reference user is blocked and out of coverage from any \ac{bs}. With this in mind, $\mathbb{P}\left(D > d \right)$ is:
\begin{eqnarray} \label{eq:P_R_correlation_2_parallel_orientation}
     && \mathbb{E}_{\Psi_{\text{BS}}}\hspace{-1mm}\left[ \sum_{\text{\textbf{X}} \in \Psi_{\text{BS}} \setminus \text{\textbf{X}} \in  \mathcal{B}\left (\text{\textbf{O}},d\right)} \hspace{-8mm} \mathbbm{1}\left(\text{closest visible BS is at \textbf{X}} \right) \right] + \mathbb{P}\left(D = \infty \right) \nonumber \\
     &&\overset{\text{(a)}}{=} \lambda_{\text{BS}} \int_{\mathbb{R}^2\setminus\mathcal{B}\left (\text{\textbf{O}},d\right)} \hspace{-10mm} \mathbb{E}^{\text{\textbf{x}}} \left[\mathbbm{1}\left(\text{closest visible BS is at \textbf{x}} \right) \right] \text{d\textbf{x}} + \mathbb{P}\left(D = \infty \right) \nonumber \\
     &&=\lambda_{\text{BS}} \int_{\mathbb{R}^2\setminus\mathcal{B}\left (\text{\textbf{O}},d\right)} \hspace{-4mm} \mathbb{P}^{\text{\textbf{x}}} \left[\text{closest visible BS is at \textbf{x}} \right] \text{d\textbf{x}} + \mathbb{P}\left(D = \infty \right), \nonumber \\
\end{eqnarray}
where (a) comes from applying the Campbell-Mecke theorem, where $\mathbb{P}^{\text{\textbf{x}}} \left[\cdot \right]$ is the Palm probability \cite{Gupta2020DoesBC,Andrews2016APO}, i.e., the probability conditioned on the fact that there is a \ac{bs} at $\mathbf{x}\in\Psi_{\text{BS}}$. Note that the probability that the reference user has not any visible \ac{bs} is taken into account through $\mathbb{P}\left(D = \infty \right)$. 

Developing \eqref{eq:P_R_correlation_2_parallel_orientation} and considering the event:
\begin{align}\label{eq:Event_J_x}
    J_{\text{\textbf{x}}} = \left\{\text{no visible BS is closer than \textbf{x}} \right\},
\end{align}
we have that:
\begin{eqnarray} \label{eq:P_closest_visible_AP_at_x_parallel_orientation}
     \hspace{-1mm} \mathbb{P}^{\text{\textbf{x}}} \hspace{-1mm} \left[\text{closest visible BS is at \textbf{x}} \right] = \mathbb{P}^{\text{\textbf{x}}} \hspace{-1mm}\left(J_{\text{\textbf{x}}}|\text{LOS}_{\text{\textbf{x}}} \right) \mathbb{P}^{\text{\textbf{x}}} \hspace{-1mm}\left(\text{LOS}_{\text{\textbf{x}}} \right)\hspace{-1mm}.
\end{eqnarray}
Then, the focus is put on the term $\mathbb{P}^{\text{\textbf{x}}} \left(J_{\text{\textbf{x}}}|\text{LOS}_{\text{\textbf{x}}} \right)$. The occurrence of $J_{\text{\textbf{x}}}$ means that all the \acp{bs} different from the \ac{bs} at \textbf{x} that are in \ac{los} (i.e., non-blocked), are further from the origin than the one in \textbf{x} (i.e., all the \acp{bs} closer than $\|\text{\textbf{x}}\|$ from the origin have to be non visible). Therefore, $\mathbb{P}^{\text{\textbf{x}}} \left(J_{\text{\textbf{x}}}|\text{LOS}_{\text{\textbf{x}}} \right)$
\begin{equation}\label{eq:P_J_x_LOS_x}
    \hspace{-3.5mm}\approx  \exp{\hspace{-1mm}\left(\hspace{-1mm}-\lambda_{\text{BS}} \hspace{-1mm} \int_0^{2\pi} \hspace{-2.5mm} \int_0^x \hspace{-0.5mm}\frac{\mathbb{P}^{\text{\textbf{x}}} \left(\text{LOS}_{\text{\textbf{x}}} \land \text{LOS}_{\text{\textbf{t}}}\big|x,\phi,t,\omega \right)}{\mathbb{P}^{\text{\textbf{x}}} \left(\text{LOS}_{\text{\textbf{x}}} \big|x,\phi\right)} t\text{d}t \text{d}\omega \hspace{-1mm} \right)}, \\
\end{equation}
where the approximation accounts for the fact that only the correlation between pairs of links is considered. The full development, extracted from \cite{Gupta2020DoesBC}, can be followed in \refappendix{subsecapp:Derivation_of_P(J_x|LOS_x)}.

If we rename
\begin{equation} \label{eq:g(x,phi)}
    g(x,\phi) = \mathbb{P}^{\text{\textbf{x}}} \left(J_{\text{\textbf{x}}}|\text{LOS}_{\text{\textbf{x}}} \right) \mathbb{P}^{\text{\textbf{x}}} \left(\text{LOS}_{\text{\textbf{x}}} \right) x,
\end{equation}
considering that $F_D(0) = 0$, and replacing $\text{\textbf{x}}=(x,\phi)$ in polar coordinates and $g(x,\phi)$ in \eqref{eq:P_R_correlation_2_parallel_orientation}, we have that:
\begin{equation} \label{eq:F_D_correlation}
    F_D(d) \approx \lambda_{\text{BS}} \int_0^{2\pi} \int_0^d g(x,\phi) \text{d}x \text{d}\phi.
\end{equation}
To obtain a closed-form expression for $g(x,\phi)$, we need to compute $\mathbb{P}^{\text{\textbf{x}}} \left(\text{LOS}_{\text{\textbf{x}}}\big| x,\phi\right)$ \\and $\mathbb{P}^{\text{\textbf{x}}} \left(\text{LOS}_{\text{\textbf{x}}} \land \text{LOS}_{\text{\textbf{t}}}\big|x,\phi,t,\omega \right)$, as it was done in \cite{Gupta2020DoesBC} for the random orientation assumption. While  $\mathbb{P}^{\text{\textbf{x}}} \left(\text{LOS}_{\text{\textbf{x}}} \big|x,\phi\right) = \exp\left(-\lambda_{\text{b}} \mathbb{E}\left[L \right] x \sin{\phi} \right)$, the computation of $\mathbb{P}^{\text{\textbf{x}}} \left(\text{LOS}_{\text{\textbf{x}}} \land \text{LOS}_{\text{\textbf{t}}}\big|x,\phi,t,\omega \right)$ is not straightforward. This probability is equivalent to the probability that no points representing the centers of blocking elements fall within the union of the blocking regions of the two links considered at the same time, $S_1\cup S_2 (l,x,\phi,t,\omega)$, of area $A_{S_1\cup S_2 (l,x,\phi,t,\omega)}$.

\subsubsection{Upper bound of the area}\label{subsubsec:Upper_bound_of_the_area}
As a first approximation, by upper bounding the area\\ $A_{S_1\cup S_2 (l,x,\phi,t,\omega)} \leq A_{S_1} + A_{S_2} = l \left(x \lvert \sin{\phi}\rvert + t \lvert \sin{\omega}\rvert \right)$, we have that $\mathbb{P}^{\text{\textbf{x}}} \left(\text{LOS}_{\text{\textbf{x}}} \land \text{LOS}_{\text{\textbf{t}}}\big|x,\phi,t,\omega \right) \geq e^{-\lambda_{\text{b}} \mathbb{E}\left[L \right] \left(x \sin{\phi} + t \lvert \sin{\omega}\rvert \right)}$ \cite{Gupta2020DoesBC}. Together with $\mathbb{P}^{\text{\textbf{x}}} \left(\text{LOS}_{\text{\textbf{x}}} \big|x,\phi\right)$, an upper bound on $g(x,\phi)$ is
\begin{eqnarray}
    \overline{g(x,\phi)} &=& \exp{\left(-4\lambda_{\text{BS}} \int_0^{\frac{\pi}{2}} \int_0^x t e^{-\lambda_{\text{b}} \mathbb{E}\left[L \right] t  \sin{\omega}} \text{d}t \text{d}\omega \right)} \times \nonumber \\
    &&e^{-\lambda_{\text{b}} \mathbb{E}\left[L \right] x \sin{\phi}} x.
\end{eqnarray}
Plugging this result, instead of $g(x,\phi)$, in \eqref{eq:F_D_correlation}, we obtain the same upper bound of $F_D(d)$ given in \eqref{eq:F_D(d)_independent} for independent blockings. This is because the bound considered for the area $A_{S_1\cup S_2}$ takes into account the overlapping area twice.

\subsubsection{Exact formulation for the area}\label{subsubsec:Exact_formulation_for_the_area}
Different from \cite{Gupta2020DoesBC}, we derive the exact formulation for $A_{S_1\cup S_2 (l,x,\phi,t,\omega)}$ (see \refappendix{subsecapp:Derivation_of_the_expression_for_the_area}), to compute $\mathbb{P}^{\text{\textbf{x}}} \left(\text{LOS}_{\text{\textbf{x}}} \land \text{LOS}_{\text{\textbf{t}}}\big|x,\phi,t,\omega \right)$. Plugging it and $\mathbb{P}^{\text{\textbf{x}}} \left(\text{LOS}_{\text{\textbf{x}}} \big|x,\phi\right)$ in \eqref{eq:P_J_x_LOS_x} and in \eqref{eq:g(x,phi)} respectively, we derive an exact expression for $g(x,\phi)$ for $0 \leq \phi \leq \frac{\pi}{2}$. Given the symmetry of the functions, the closed form expression for $g\left(x,\phi \right)$ for other values of $\phi$ greater than $\frac{\pi}{2}$ can be obtained following the same considerations. Finally, similarly to \autoref{subsubsec:Upper_bound_of_the_area}, we obtain the lower bound of $F_D(d)$ by replacing $g\left(x,\phi \right)$ in \eqref{eq:F_D_correlation}.

\vspace{-0.25cm}
\section{Inclusion of the Height} \label{sec:Inclusion_Height}
\begin{figure}[t]
	\centering
	\includegraphics[width=7cm]{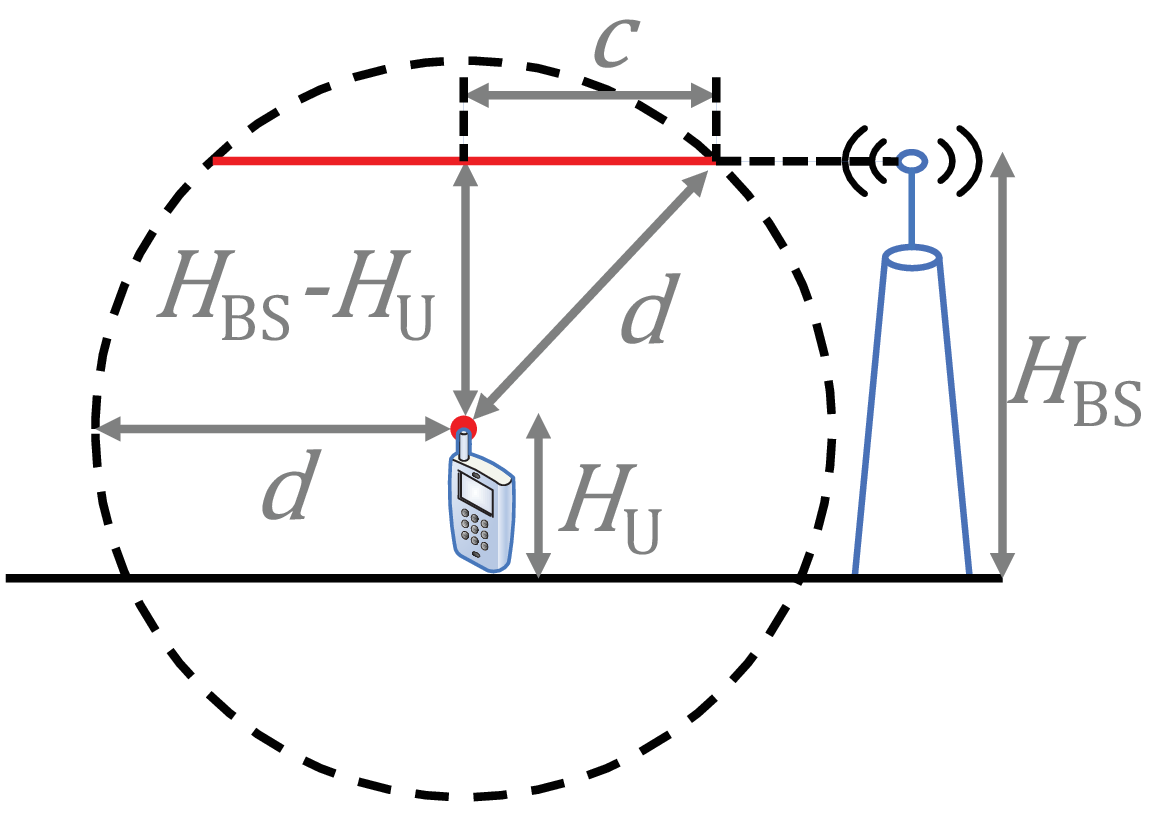}
	\caption{Representation of a vertical cut of the ball $\mathcal{B}\left (\text{\textbf{O}},d\right)$ in $\mathbb{R}^3$ and its associated disk of radius $c$ in $\mathbb{R}^2$.}\vspace{-0.4cm}
	\label{fig:Heights_distances_ball}
\end{figure}
In this section, we present the methodology to extend the previous derivations to account for the heights of the blockages, user, and \acp{bs} (denoted by $h_{\text{b}}$, $H_{\text{U}}$, and $H_{\text{BS}}$, respectively) as shown in \autoref{fig:Heights_distances_ball}. For in-street users, this extension is not needed as users would be lower than all the buildings, which implies that any building in the line connecting a user and a \ac{bs} would block the transmission.

In order to obtain the \ac{los} probabilities, a so-called scaling factor $\eta$ needs to be considered as in \cite{bai2013,garcia2021}, which accounts for the fraction of buildings that effectively block the link when considering the heights. If we denote by $c$ the projection of the distance $d$ onto the 2-dimensional plane, it holds that \\ $c(d) = \sqrt{d^2-{\left(H_{\text{BS}}-H_{\text{U}}\right)}^2}$. Consequently, the minimum value for $d$ will be $d_{\min} = H_{\text{BS}}-H_{\text{U}}$.

In \autoref{sec:Distribution_Distance_Closest_Visible_BS}, we have obtained the probability that a \ac{bs} falls within a ball based on different assumptions. In this case, the same concept applies, although this ball $\mathcal{B}\left (\text{\textbf{O}},d\right)$ is in $\mathbb{R}^3$ as seen in the vertical cut of \autoref{fig:Heights_distances_ball}. However, since these points have a fixed height $H_{\text{BS}}$, $\Psi_{\text{BS}}$ is indeed a \ac{ppp} in $\mathbb{R}^2$. Then, the formulation of the probability that $m$ points fall within the ball $\mathcal{B}\left (\text{\textbf{O}},d\right)$ in $\mathbb{R}^3$ is equivalent to the probability of $m$ points falling within its associated disk in $\mathbb{R}^2$ of radius $c(d)$, that is, the ball $\mathcal{B}\left (\text{\textbf{O}},c(d)\right)$ in red in \autoref{fig:Heights_distances_ball}. Then, the procedure is the same as the one followed in \autoref{sec:Distribution_Distance_Closest_Visible_BS}.

\vspace{-0.25cm}
\section{Results} \label{sec:Results_2D_parallel_orientation}
\subsection{Validation of the Analytical Bounds of the \acp{cdf}}\label{subsec:Bounds_for_the_CDFs}
\begin{figure}[t]
	\centering
	\includegraphics[width=10cm]{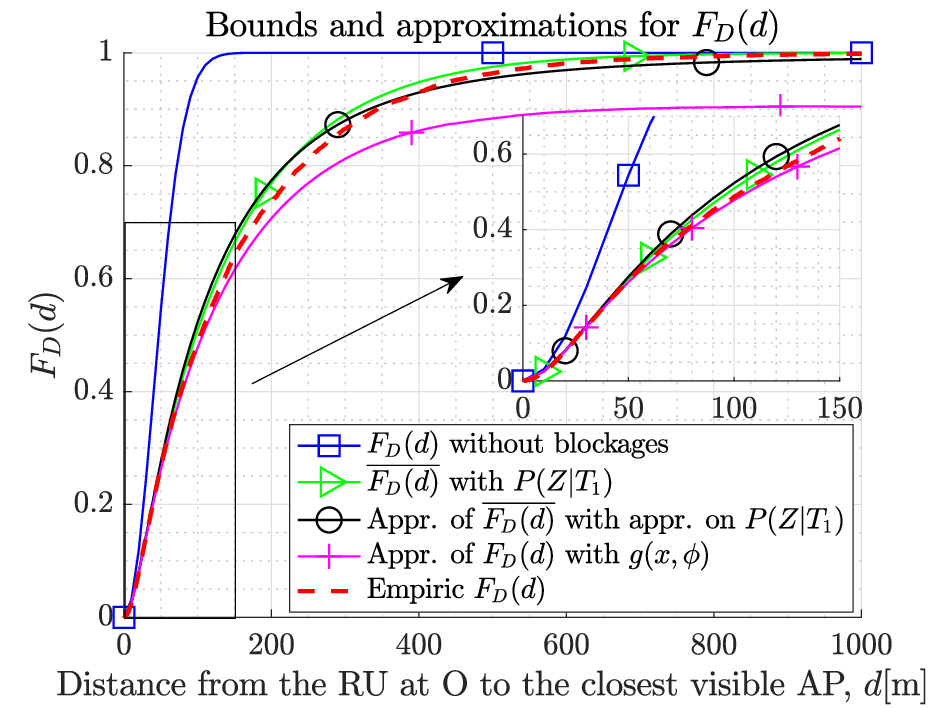}
			\caption{Empirical $F_D(d)$ and comparison with the derived analytical bounds and the case of no blockages.}
	\label{fig:CDFs_and_PDFs_parallel_orientation}

\end{figure}
\begin{figure}[t]
	\centering
	\includegraphics[width=10cm]{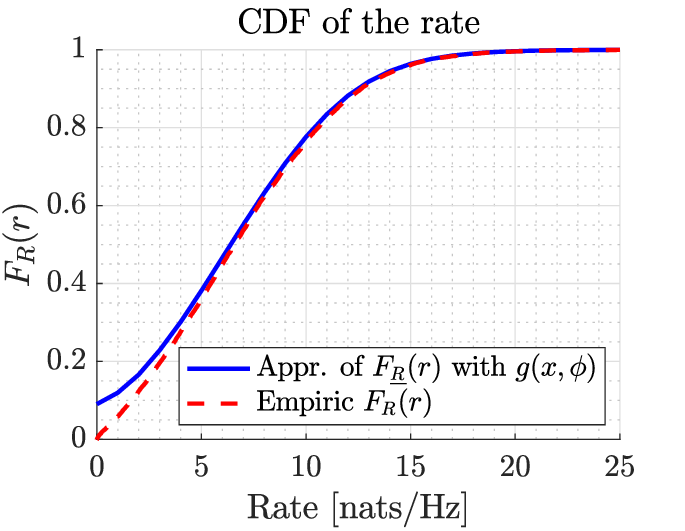}
		\caption{Empirical $F_R(r)$ and comparison with the derived analytical bound.}
	\vspace{-0.4cm}
	\label{fig:Rate_CDF_parallel_orientation}
\end{figure}
\begin{figure}[t]
	\centering
	\includegraphics[width=10cm]{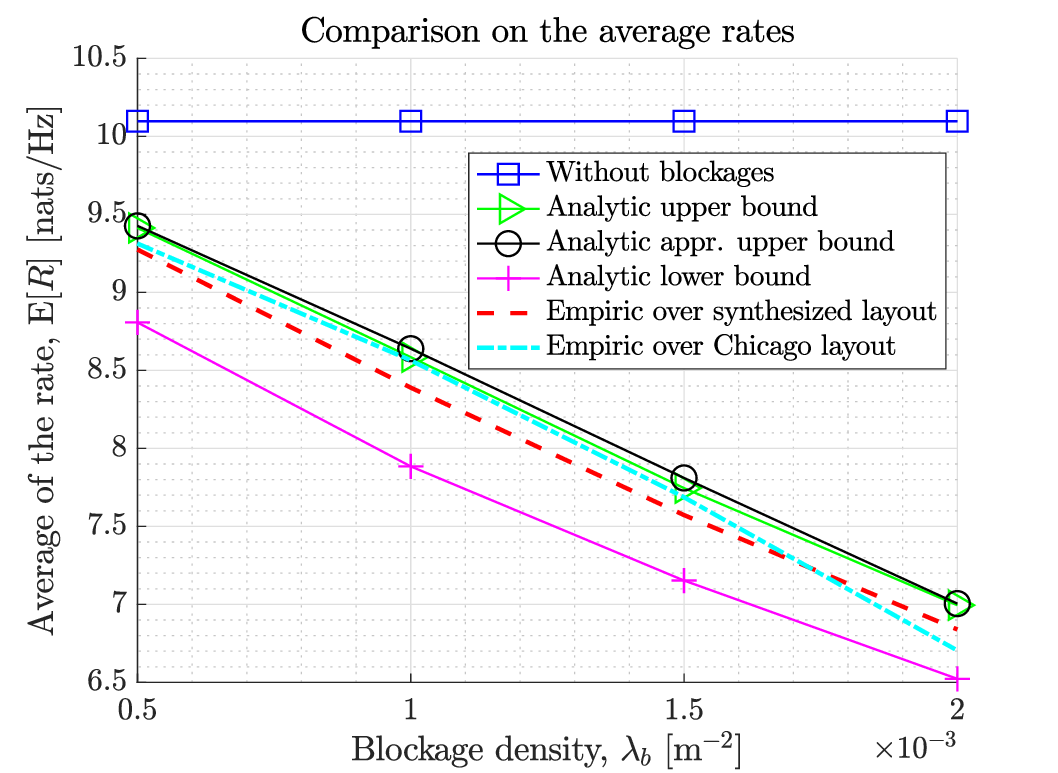}
	\caption{Comparison of the average rates (both analytical and empirical results) for different layouts.}
	\vspace{-0.3cm}
	\label{fig:Average_Rates}
\end{figure}
In this section, we validate the obtained analytical expressions by comparing them with empirical results over generated layouts with blockages and \acp{bs} following a \ac{ppp} distribution according to \autoref{sec:System_Model}. The simulation parameters are: $\lambda_{\text{BS}}=1\cdot 10^{-4} \hspace{1mm}\SI{}{BSs/m^2}$, $\lambda_{\text{b}}=1.9\cdot 10^{-3} \hspace{1mm}\SI{}{blockings/m^2}$, and $L$ being uniformly distributed between $\SI{0}{\meter}$ and $\SI{57}{\meter}$.

\autoref{fig:CDFs_and_PDFs_parallel_orientation} shows the different derived analytical bounds and approximations for the \ac{cdf} of $D$, $F_D(d)$. The approximation of $F_D(d)$ with $g(x,\phi)$ acts as a lower bound because of the approximation of the correlation by pairs of links. Note that, for lower values of $d$, the empirical results are close to the approximation where the one-by-one correlation is considered, while for greater values of $d$, it is closer to the upper bound derived based on the independence assumption.

On the other hand, we take the approximation of the \ac{cdf} of the distance with $g\left(x,\phi\right)$ to approximate $F_{\underline{R}}(r)$. The analytical expression and empirical simulations can be compared in \autoref{fig:Rate_CDF_parallel_orientation} for $P_{\text{U}}= \SI{33}{dBm}$, $\sigma^2 = \SI{-104}{dBm}$, $L_{\text{ref}} = \SI{25.6}{dB}$, and $\alpha = 4$. We see that this analytical expression follows quite accurately the empirical curve.

\vspace{-0.2cm}
\subsection{Model Assessment}\label{subsec:Bounds_for_the_average_rate}

To assess the validity of our model, we present in \autoref{fig:Average_Rates} the average rate obtained by numerical simulations for an actual building layout in Chicago \cite{measurementsKulkarni} and synthetic layouts generated as described in Section II, and also the rate computed from the derived bounds. To evaluate rate versus $\lambda_b$ in the Chicago deployment, we randomly take subsets of buildings and count each building as two pairs of parallel lines. For the results to be comparable, an empirical correction factor of 1.7  was still required for the density of the lines in our synthetic model, as having two orientations for the blockings in the actual layout instead of one increases the blockage probability. Also, note that modeling the length of the blocking elements as a uniform r.v. prevents having high length values, although this happens in reality with low probability. In any case, after applying the constant correction factor, the results are very adjusted for all the blocking densities.

\section{Conclusions}\label{sec:Conclusions}
In this work, we have obtained an approximation of the \ac{cdf} of the distance to the closest visible \ac{bs} in urban scenarios. Modeling the blockages as parallel line segments, we have provided bounds of this \ac{cdf} for independent and correlated blockages and the ergodic capacity. We have also presented a method to consider the height of the obstacles and compared the synthetic system model with a real layout of buildings.

Future work could extend the analytical derivation to other building models more sophisticated than those based on line segments in order to represent reality more accurately. Also, the rate analysis could incorporate effects produced by human blocking and atmospheric effects. Finally, heterogeneous networks with different tiers of \acp{bs} could be considered.

\vspace{-0.3cm}
\appendix\label{sec:Appendix}
\vspace{-0.3cm}
\subsection{Derivation of $\underline{\mathbb{P}\left(D > d \right)}$}\label{subsecapp:Derivation_of_P_D(d)_lb} 
From the definition of $T_p$ in \eqref{eq:Event_Tp} and that $P \sim \mathcal{P}\left(\lambda_{\text{BS}} \pi d^2 \right)$:
\begin{eqnarray}
    \underline{\mathbb{P}\left(D > d \right)} &=& \underline{\mathbb{P}\left(Z \right)} = \sum_{p=0}^{\infty} \underline{\mathbb{P}\left(Z | T_p\right)} \mathbb{P}\left(T_p \right) \nonumber \\
    &=& \sum_{p=0}^{\infty} {\mathbb{P}\left(Z | T_1 \right)}^p \frac{{\left( \lambda_{\text{BS}} \pi d^2 \right)}^p}{p!}e^{-\lambda_{\text{BS}} \pi d^2} \nonumber \\
    &\overset{\text{(a)}}{=}& e^{-\lambda_{\text{BS}} \pi d^2 \left(1-\mathbb{P}\left(Z | T_1 \right)  \right)},
\end{eqnarray}
where the lower bounds on $\underline{\mathbb{P}\left(D > d \right)}$, $\underline{\mathbb{P}\left(Z \right)}$, and $\underline{\mathbb{P}\left(Z | T_p\right)}$ come from the independence assumption and (a) comes from the fact that $\sum_{p=0}^{\infty} = \frac{y^p}{p!} = e^{y}$ \cite{bai2013}.

\subsection{Derivation of $\mathbb{P}^{\text{\textbf{x}}} \left(J_{\text{\textbf{x}}}|\text{LOS}_{\text{\textbf{x}}} \right)$}\label{subsecapp:Derivation_of_P(J_x|LOS_x)}
Regarding the definition of $J_{\text{\textbf{x}}}$ in \eqref{eq:Event_J_x}, we have that:
$\mathbb{P}^{\text{\textbf{x}}} \left(J_{\text{\textbf{x}}}|\text{LOS}_{\text{\textbf{x}}} \right) =$
\begin{eqnarray} \label{eq:P_J_x_LOS_x_development}
    &=& \mathbb{P}^{\text{\textbf{x}}} \left(\bigwedge\limits_{\text{\textbf{t}}\in\Psi_{\text{BS}}} \text{BS in \textbf{t} is closer than \textbf{x} but in \text{NLOS}} \Big|\text{LOS}_{\text{\textbf{x}}} \right) \nonumber \\
    &=& \mathbb{P}^{\text{\textbf{x}}} \left(\bigwedge\limits_{\text{\textbf{t}}\in\Psi_{\text{BS}}} \|\text{\textbf{t}}\|\leq\|\text{\textbf{x}}\| \land \text{NLOS}_{\text{\textbf{t}}}  \Big|\text{LOS}_{\text{\textbf{x}}} \right) \nonumber \\
    &\overset{\text{(a)}}{\approx}& \mathbb{P}^{\text{\textbf{x}}} \left(\prod_{\text{\textbf{t}}\in\Psi_{\text{BS}}} \mathbbm{1}\left(\|\text{\textbf{t}}\|\leq\|\text{\textbf{x}}\| \right) \mathbbm{1}\left(\text{NLOS}_{\text{\textbf{t}}} \right)  \Big|\text{LOS}_{\text{\textbf{x}}} \right) \nonumber \\
    &=& \mathbb{P}^{\text{\textbf{x}}} \left(\prod_{\substack{\text{\textbf{t}}\in\Psi_{\text{BS}}\cap \mathcal{B}\left (\text{\textbf{O}},\|\text{\textbf{x}}\|\right)}} \mathbbm{1}\left(\text{NLOS}_{\text{\textbf{t}}} \right)  \Big|\text{LOS}_{\text{\textbf{x}}} \right) \nonumber \\
    &\overset{\text{(b)}}{=}& \exp{\left(-\lambda_{\text{BS}} \int_{\mathcal{B}\left (\text{\textbf{O}},\|\text{\textbf{x}}\|\right)} \left(1 - \mathbb{P}^{\text{\textbf{x}}} \left(\text{NLOS}_{\text{\textbf{t}}} \big|\text{LOS}_{\text{\textbf{x}}}\right) \right) \text{d\textbf{t}} \right)} \nonumber \\
    &\overset{\text{(c)}}{=}& \exp{\left(-\lambda_{\text{BS}}\int_0^{2\pi}\int_0^x \mathbb{P}^{\text{\textbf{x}}} \left(\text{LOS}_{\text{\textbf{t}}}  \big|\text{LOS}_{\text{\textbf{x}}}, x,\phi,t,\omega \right) t\text{d}t \text{d}\omega \right)} \nonumber \\
    &\overset{\text{(d)}}{=}&\exp{\left(-\lambda_{\text{BS}}\int_0^{2\pi}\int_0^x \frac{\mathbb{P}^{\text{\textbf{x}}} \left(\text{LOS}_{\text{\textbf{x}}} \land \text{LOS}_{\text{\textbf{t}}}\big|x,\phi,t,\omega \right)}{\mathbb{P}^{\text{\textbf{x}}} \left(\text{LOS}_{\text{\textbf{x}}} \big|x,\phi\right)} t\text{d}t \text{d}\omega \right)},\nonumber \\
\end{eqnarray}
where (a) is an approximation given that only the correlation between the reference link and another link is considered in a one-by-one basis, (b) comes from the \ac{pgfl} of a \ac{ppp} (see (15) in \cite{Andrews2016APO}, \cite{2016DiRenzoIntro} and Definition 2.4. in \cite{2008KrishnaL9}), (c) comes from the definitions $\text{\textbf{x}}=(x,\phi)$ and $\text{\textbf{t}}=(t,\omega)$ in polar coordinates, and (d) comes from Bayes' formula for the conditional probability \cite{Gupta2020DoesBC}.

\subsection{Derivation of $A_{S_1\cup S_2 (l,x,\phi,t,\omega)}$}\label{subsecapp:Derivation_of_the_expression_for_the_area}
Consider one user and two \acp{bs}, that is, two links. Similar to \autoref{fig:Blocking_region_single}, \autoref{fig:Blocking_regions_t_omega} depicts the regions $S_1$ and $S_2$ representing the geometric locus of the centers of the line segments that block link 1 (reference link) and link 2, respectively. Note that, as we are not considering heights, both links are in the same plane. The variables $x$ and $\phi$ represent the length and azimuth  of link 1, while $t$ and $\omega$ do it for link 2. Finally, $s$ is an auxiliary variable whose value is defined in relation to  $t$. 

 As we are interested in the probability of having both links in \ac{los}, no blockings elements must have a center within the union of   $S_1$ and $S_2$. To derive a closed form expression for $A_{S_1\cup S_2 (l,x,\phi,t,\omega)}$, we consider first the case $0 \leq \phi \leq \frac{\pi}{2}$. In such case, whenever $\pi \leq \omega \leq 2\pi$, there is no overlapping between the blocking regions $S_1$ and $S_2$, leading to 
\vspace{-0.1 cm}
\begin{equation}
A_{S_1\cup S_2 (l,x,\phi,t,\omega)} = A_{S_1} + A_{S_2} = l \left(x \lvert \sin{\phi}\rvert + t \lvert \sin{\omega}\rvert \right).    
\end{equation}

The difficulty arises when the overlapping is not null, which happens for $0 \leq \omega \leq \pi$. In \autoref{fig:Blocking_regions_t_omega} four different margins of values for both $t$ and $\omega$ are considered:
\begin{figure}[t]
	\centering
	\includegraphics[width=10 cm]{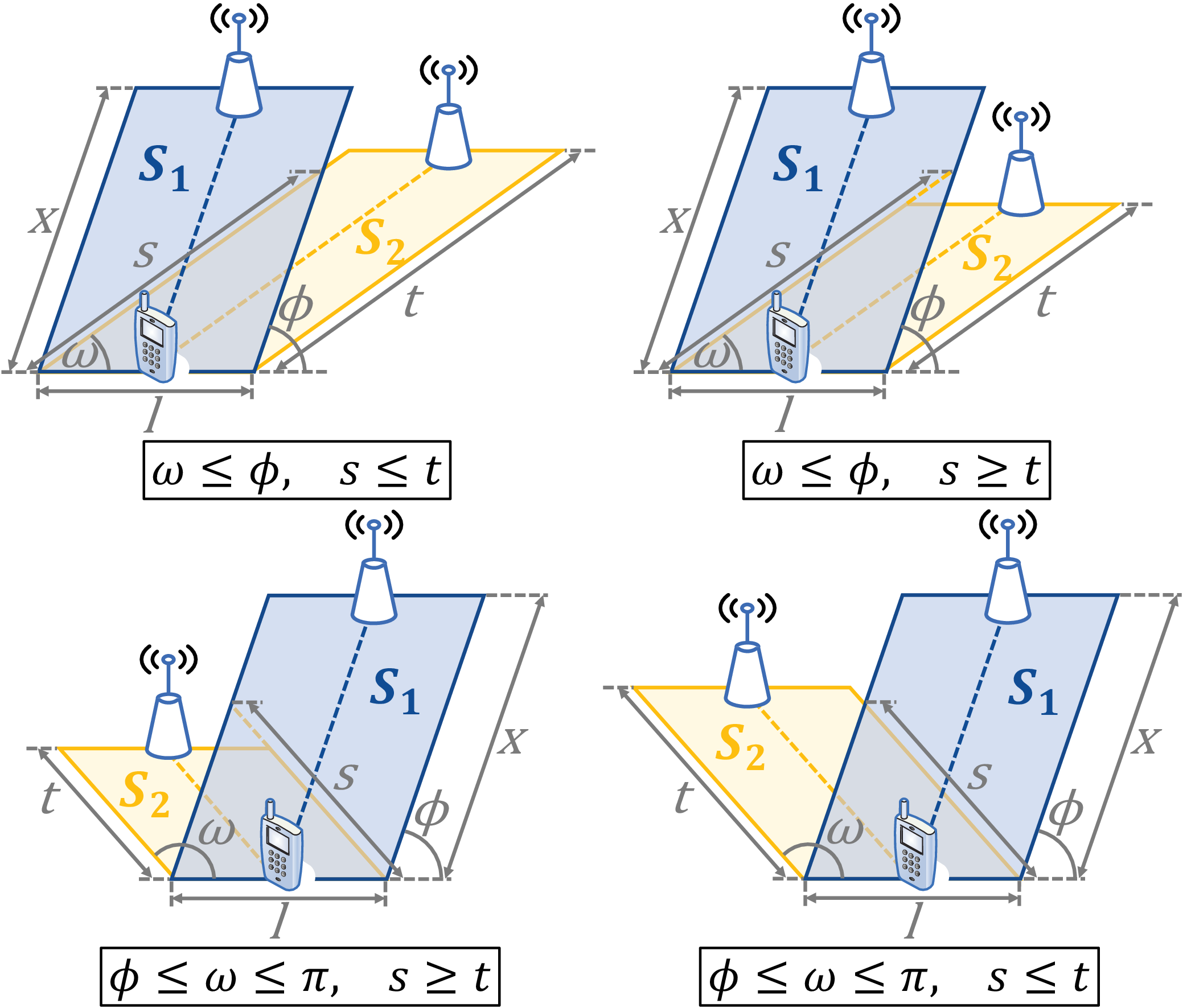}
	\caption{Blocking regions $S_1$, $S_2$ for a given length $l$, distance $x$, azimuth $\phi$, and different distances $t$ and azimuths $\omega$.}\vspace{-0.4cm}
	\label{fig:Blocking_regions_t_omega}
\end{figure}

\begin{itemize}
    \item for $\omega \leq \phi$ and $s = l \frac{\sin\phi}{\sin{\left(\phi - \omega \right)}} \leq t$  we have that $A_{S_1\cup S_2 (l,x,\phi,t,\omega)} = l x \sin{\phi} + l t \sin{\omega} - l^2 \frac{\sin\phi \sin\omega}{2\sin{\left(\phi - \omega \right)}}$,
    \item for $\omega \leq \phi$ and $s = l \frac{\sin\phi}{\sin{\left(\phi - \omega \right)}} \geq t$ we have that $A_{S_1\cup S_2 (l,x,\phi,t,\omega)} = l x \sin{\phi} + t^2 \frac{\sin\omega \sin{\left(\phi - \omega \right)}}{2\sin\phi}$,
    \item for $\phi \leq \omega \leq \pi$ and $s = l \frac{\sin\phi}{\sin{\left(\omega - \phi \right)}} \geq t$ we have that $A_{S_1\cup S_2 (l,x,\phi,t,\omega)} = l x \sin{\phi} + t^2 \frac{\sin\omega \sin{\left(\omega - \phi \right)}}{2\sin\phi}$,
    \item for $\phi \leq \omega \leq \pi$ and $s = l \frac{\sin\phi}{\sin{\left(\omega - \phi \right)}} \leq t$ we have that $A_{S_1\cup S_2 (l,x,\phi,t,\omega)} = l x \sin{\phi} + l t \sin{\omega} - l^2 \frac{\sin\phi \sin\omega}{2\sin{\left(\omega - \phi \right)}}$.
\end{itemize}

To sum up the formulation for $0 \leq \phi \leq \frac{\pi}{2}$, also including the case for $\pi \leq \omega \leq 2\pi$ where there is no overlapping, is the following:

\begin{equation}\label{eq:A_S_1_U_S_2_l_parallel_orientation}
    \hspace{-5.9cm}A_{S_1\cup S_2 (l,x,\phi,t,\omega)} =
\end{equation}
\begin{equation}
		\left\{ 
			\begin{array}{l}
         		\l x \sin{\phi} + l t \sin{\omega} - l^2 \frac{\sin\phi \sin\omega}{2\lvert{\sin{\left(\phi - \omega \right)}}\rvert} \\ \hspace{2.8cm} \text{for} \hspace{0.3cm} 0 \leq \omega \leq \pi\text{, }0 \leq l \leq t\frac{\lvert{\sin{\left(\phi - \omega \right)}}\rvert}{\sin\phi}; \\
             	l x \sin{\phi} + t^2 \frac{\sin\omega\lvert{\sin{\left(\phi - \omega \right)}}\rvert}{2\sin\phi} \\ \hspace{2.8cm} \text{for} \hspace{0.3cm} 0 \leq \omega \leq \pi\text{, } t\frac{\lvert{\sin{\left(\phi - \omega \right)}}\rvert}{\sin\phi} \leq l \leq \infty; \\
             	l \left(x \sin{\phi} + t \lvert \sin{\omega}\rvert \right)   \text{ for}  \hspace{0.3cm} \pi \leq \omega \leq 2\pi.	\end{array}\nonumber
   		\right.
\end{equation}

Consequently, for $0 \leq \phi \leq \frac{\pi}{2}$, we have that $\mathbb{E}\left[K_{1,2} | x,\phi,t,\omega \right] = \lambda_{\text{b}} \mathbb{E}\left[L \right] \left(x \sin{\phi} + t \lvert \sin{\omega}\rvert \right)$ whenever $\pi \leq \omega \leq 2\pi$. For other values of $0 \leq \omega \leq \pi$, let us define the function $a \coloneqq a(\phi,t,\omega) = t\frac{\lvert{\sin{\left(\phi - \omega \right)}}\rvert}{\sin\phi}$ for notation purposes. As it is highlighted in \eqref{eq:A_S_1_U_S_2_l_parallel_orientation}, for those values of the azimuth $\omega$, the expression for the area depends on $l$ being smaller or greater than $a$. At this point, to obtain a closed form expression for $\mathbb{E}\left[K_{1,2} | x,\phi,t,\omega \right]$ for these values of azimuths and distances to the \acp{bs}, the statistics of $L$ need to be considered. In this case, we assume $L \sim \mathcal{U}\left[L_{\min}, L_{\max}\right]$, with $\mathcal{U}$ denoting a uniform distribution between $L_{\min}$ and $L_{\max}$. Therefore, if $0 \leq \phi \leq \frac{\pi}{2}$ and $0 \leq \omega \leq \pi$, we will have different expressions for $\mathbb{E}\left[K_{1,2} | x,\phi,t,\omega \right]$, depending on whether $0 \leq a \leq L_{\min}$, $L_{\min} \leq a \leq L_{\max}$, or $L_{\max} \leq a$.

Gathering all the expressions for $0 \leq \phi \leq \frac{\pi}{2}$, we obtain \eqref{eq:E_K_1_2_l_parallel_orientation}, and with that \\$\mathbb{P}^{\text{\textbf{x}}} \left(\text{LOS}_{\text{\textbf{x}}} \land \text{LOS}_{\text{\textbf{t}}}\big|x,\phi,t,\omega \right) = \exp\left(-\mathbb{E}\left[K_{1,2} | x,\phi,t,\omega \right] \right)$.

\begin{equation}\label{eq:E_K_1_2_l_parallel_orientation}
    \hspace{-5.9cm}\mathbb{E}\left[K_{1,2} | x,\phi,t,\omega \right]=
\end{equation}

\begin{equation}
	\left\{ 
	       \begin{array}{ll}
		 \lambda_{\text{b}} \left(\mathbb{E}\left[L \right]  x \sin{\phi} + t^2 \frac{\sin\omega \lvert{\sin{\left(\phi - \omega \right)}}\rvert}{2\sin\phi} \right) \\&\hspace{-5.5cm}\text{ for } 0 \leq \omega \leq \pi\text{, } 0 \leq a \leq L_{\min}; \\
		 \lambda_{\text{b}}\Big( \mathbb{E}\left[L \right] x \sin{\phi} + \frac{a^2-{L_{\min}}^2}{2\left(L_{\max}-L_{\min} \right)} t \sin{\omega} \\
		 \qquad - \frac{a^3-{L_{\min}}^3}{6\left(L_{\max}-L_{\min} \right)} \frac{\sin\phi \sin\omega}{\lvert{\sin{\left(\phi - \omega \right)}}\rvert} 
		 + \frac{L_{\max} - a}{L_{\max}-L_{\min}} t^2 \frac{\sin\omega \lvert{\sin{\left(\phi - \omega \right)}}\rvert}{2\sin\phi} \Big)\\ &\hspace{-5.5cm}\text{ for } 0 \leq \omega \leq \pi\text{, } L_{\min} \leq a \leq L_{\max}; \\
		 \lambda_{\text{b}} \Big(\mathbb{E}\left[L \right]  x \sin{\phi} + \mathbb{E}\left[L \right]  t \sin{\omega} 
		 - \frac{{L_{\max}}^3-{L_{\min}}^3}{6\left(L_{\max}-L_{\min}\right)} \frac{\sin\phi \sin\omega}{\lvert{\sin{\left(\phi - \omega \right)}}\rvert} \Big)\\ &\hspace{-5.5cm}\text{ for } 0 \leq \omega \leq \pi\text{, } L_{\max} \leq a;\nonumber \\
		 \lambda_{\text{b}} \mathbb{E}\left[L \right] \left(x \sin{\phi} + t \lvert \sin{\omega}\rvert \right) &\hspace{-5.3cm}\text{ for }  \pi \leq \omega \leq 2\pi.
		\end{array}
   		\right.
\end{equation}

\bibliography{blocking_references}

\begin{thebibliography}{10}
\providecommand{\url}[1]{#1}
\csname url@samestyle\endcsname
\providecommand{\newblock}{\relax}
\providecommand{\bibinfo}[2]{#2}
\providecommand{\BIBentrySTDinterwordspacing}{\spaceskip=0pt\relax}
\providecommand{\BIBentryALTinterwordstretchfactor}{4}
\providecommand{\BIBentryALTinterwordspacing}{\spaceskip=\fontdimen2\font plus
\BIBentryALTinterwordstretchfactor\fontdimen3\font minus \fontdimen4\font\relax}
\providecommand{\BIBforeignlanguage}[2]{{%
\expandafter\ifx\csname l@#1\endcsname\relax
\typeout{** WARNING: IEEEtran.bst: No hyphenation pattern has been}%
\typeout{** loaded for the language `#1'. Using the pattern for}%
\typeout{** the default language instead.}%
\else
\language=\csname l@#1\endcsname
\fi
#2}}
\providecommand{\BIBdecl}{\relax}
\BIBdecl

\bibitem{bai2013}
T.~Bai, R.~Vaze, and R.~W. Heath, ``Analysis of blockage effects on urban cellular networks,'' \emph{IEEE Trans. on Wireless Communications}, vol.~13, no.~9, pp. 5070--5083, Sep. 2014.

\bibitem{hriba:2019}
E.~Hriba and M.~Valenti, ``Correlated blocking in {mmWave} cellular networks: Macrodiversity, outage, and interference,'' \emph{Electronics}, vol.~8, no. 1187, pp. 1--17, Oct. 2019.

\bibitem{Gupta2020DoesBC}
S.~Gupta and A.~K. Gupta, ``Does blockage correlation matter in the performance of mmwave cellular networks?'' \emph{2020 IEEE Global Communications Conference (GLOBECOM)}, pp. 1--6, Dec. 2020.

\bibitem{garcia2021}
C.~G. {Ruiz}, A.~{Pascual-Iserte}, and O.~{Muñoz}, ``Analysis of blocking in mmwave cellular systems: Application to relay positioning,'' \emph{IEEE Trans. on Communications}, vol.~69, no.~2, pp. 1329--1342, Feb. 2021.

\bibitem{Article_correlation}
A.~K. Gupta, J.~G. Andrews, and R.~W. Heath, ``Macrodiversity in cellular networks with random blockages,'' \emph{IEEE Trans. on Wireless Communications}, vol.~17, no.~2, pp. 996--1010, Feb. 2018.

\bibitem{garcia2020}
C.~G. {Ruiz}, A.~{Pascual-Iserte}, and O.~{Muñoz}, ``Analysis of blocking in mmwave cellular systems: Characterization of the {LOS} and {NLOS} intervals in urban scenarios,'' \emph{IEEE Trans. on Vehicular Technology}, vol.~69, no.~12, pp. 16\,247--16\,252, Dec. 2020.

\bibitem{ITU-R_P.1410-5}
``{ITU-R P.1410-5: Propagation data and prediction methods required for the design of terrestrial broadband radio access systems operating in a frequency range from 3 to 60 GHz},'' ITU-R, Tech. Rep., Febr. 2012.

\bibitem{Samuylov2016}
A.~Samuylov, M.~Gapeyenko, D.~Moltchanov, M.~Gerasimenko, S.~Singh, N.~Himayat, S.~Andreev, and Y.~Koucheryavy, ``Characterizing spatial correlation of blockage statistics in urban mmwave systems,'' in \emph{2016 IEEE Globecom Workshops (GC Wkshps)}, Dec. 2016, pp. 1--7.

\bibitem{jain2019TIM}
I.~K. Jain, R.~Kumar, and S.~S. Panwar, ``The impact of mobile blockers on millimeter wave cellular systems,'' \emph{IEEE Journal on Selected Areas in Communications}, vol.~37, no.~4, pp. 854--868, Feb. 2019.

\bibitem{Conference_correlation}
A.~K. Gupta, J.~G. Andrews, and R.~W. Heath, ``Impact of correlation between link blockages on macro-diversity gains in {mmWave} networks,'' in \emph{IEEE International Conference on Communications Workshops (ICC Workshops)}, Kansas City, MO (USA), May 2018, pp. 1--6.

\bibitem{calvo2012}
E.~Calvo, O.~Muñoz, J.~Vidal, and A.~Agustín, ``Downlink coordinated radio resource management in cellular networks with partial {CSI},'' \emph{IEEE Trans. on Signal Processing}, vol.~60, no.~3, pp. 1420--1431, Mar. 2012.

\bibitem{aravanis2016}
A.~I. {Aravanis}, O.~{Muñoz}, A.~{Pascual-Iserte}, and J.~{Vidal}, ``Analysis of downlink and uplink decoupling in dense cellular networks,'' in \emph{2016 IEEE 21st Int. Workshop on Computer Aided Modelling and Design of Communication Links and Networks (CAMAD)}, Oct. 2016, pp. 219--224.

\bibitem{randomshape}
T.~Bai, R.~Vaze, and R.~W. Heath, ``Using random shape theory to model blockage in random cellular networks,'' in \emph{IEEE Int. Conf. on Signal Process. and Comm. (SPCOM)}, Bangalore (India), July 2012, pp. 1--5.

\bibitem{measurementsKulkarni}
S.~Singh, M.~N. Kulkarni, A.~Ghosh, and J.~G. Andrews, ``Tractable model for rate in self-backhauled millimeter wave cellular networks,'' \emph{IEEE Journal on Selected Areas in Communications}, vol.~33, no.~10, pp. 2196--2211, Oct. 2015, {MATLAB} files available at \url{https://goo.gl/Ie39k7}.

\bibitem{calvo2009}
E.~Calvo, J.~Vidal, and J.~R. Fonollosa, ``Optimal resource allocation in relay-assisted cellular networks with partial csi,'' \emph{IEEE Transactions on Signal Processing}, vol.~57, no.~7, pp. 2809--2823, 2009.

\bibitem{Andrews2016APO}
J.~G. Andrews, A.~K. Gupta, and H.~S. Dhillon, ``A primer on cellular network analysis using stochastic geometry,'' ArXiv 1604.03183, 2016.

\bibitem{2016DiRenzoIntro}
\BIBentryALTinterwordspacing
M.~D. Renzo, ``Lecture notes in intro to stochastic geometry \& point processes,'' September 2016. [Online]. Available: \url{https://people.maths.bris.ac.uk/~macpd/sen/files/course2016/lec5.pdf}
\BIBentrySTDinterwordspacing

\bibitem{2008KrishnaL9}
\BIBentryALTinterwordspacing
R.~Krishna, ``Lecture notes in advanced topics in random wireless networks,'' September 2008. [Online]. Available: \url{https://www3.nd.edu/~mhaenggi/ee87021/summary-sep-22.pdf}
\BIBentrySTDinterwordspacing

\end{thebibliography}
\bibliographystyle{IEEEtran} 
\end{document}